\documentclass[a4paper,fleqn]{cas-dc}
\usepackage{float}
\usepackage[numbers]{natbib}
\usepackage{subfigure} 
\def\tsc#1{\csdef{#1}{\textsc{\lowercase{#1}}\xspace}}
\tsc{WGM}
\tsc{QE}
\tsc{EP}
\tsc{PMS}
\tsc{BEC}
\tsc{DE}

\begin{document}
\let\printorcid\relax
\let\WriteBookmarks\relax
\def\floatpagepagefraction{1}
\def\textpagefraction{.001}
\shorttitle{Leveraging social media news}
\shortauthors{Weiming Hu et~al.}

\title[mode = title]{A Comparative Study of Gastric Histopathology Sub-size Image
Classification: from Linear Regression to Visual Transformer }  

\author[a]{Weiming Hu}[type=editor,style=chinese,auid=000,bioid=1,]

\author[a]{Haoyuan Chen}
\author[a]{Wanli Liu}
\author[b]{Xiaoyan Li}[style=chinese]
\author[c]{Hongzan Sun}
\author[d]{Xinyu Huang}
\author[d]{Marcin Grzegorzek}
\author[a]{Chen Li}[style=chinese]
\cormark[1]
\ead{lichen201096@hotmail.com}

\address[a]{Microscopic Image and Medical Image Analysis Group, College of Medicine and Biological lnformation Engineering
, Northeastern University, 110169, Shenyang, PR China}
\address[b]{Department of Pathology, Cancer Hospital, China Medical University, Liaoning Cancer Hospital and Institute,
Shenyang 110042, PR China}
\address[c]{Department of Radiology, Shengjing Hospital, China Medical University, Shenyang, 110122, China}
\address[d]{Institute of Medical Informatics, University of Luebeck, Luebeck, Germany}
\cortext[cor1]{Corresponding author}

\begin{abstract}
Background and objective:
Gastric cancer is the fifth most common cancer in the world. At the same time, it is also the fourth most deadly cancer. Early detection of cancer exists as a guide for the treatment of gastric cancer. Nowadays, computer technology has advanced rapidly to assist physicians in the diagnosis of pathological pictures of gastric cancer. Ensemble learning is a way to improve the accuracy of algorithms, and finding multiple learning models with complementarity types is the basis of ensemble learning.
\\
Methods:
The complementarity of sub-size pathology image classifiers when machine performance is insufficient is explored in this experimental platform. We choose seven classical machine learning classifiers and four deep learning classifiers for classification experiments on the GasHisSDB database. Among them, classical machine learning algorithms extract five different image virtual features to match multiple classifier algorithms.For deep learning, we choose three convolutional neural network classifiers. In addition, we also choose a novel Transformer-based classifier.
\\
Results:
The experimental platform, in which a large number of classical machine learning and deep learning methods are performed, demonstrates that there are differences in the performance of different classifiers on GasHisSDB. Classical machine learning models exist for classifiers that classify Abnormal categories very well, while classifiers that excel in classifying Normal categories also exist. Deep learning models also exist with multiple models that can be complementarity.
\\
Conclusions:
Suitable classifiers are selected for ensemble learning, when machine performance is insufficient. This experimental platform demonstrates that multiple classifiers are indeed complementarity and can improve the efficiency of ensemble learning. This can better assist doctors in diagnosis, improve the detection of gastric cancer, and increase the cure rate.
\end{abstract}

\begin{keywords}
Gastric Histopathology\sep
Sub-size Image\sep
Robustness comparison\sep
Algorithmic complementarity\sep
Image classification\sep
\end{keywords}

\maketitle

\section{Introduction}
Gastric cancer is a serious threat to human health as a global killer disease.
According to the most recent Global Cancer Statistics Report, gastric cancer has become the fifth most common cancer and the fourth leading cause of death~\cite{sung2021global}.
Histopathological examination of gastric cancer constitutes the gold standard for the detection of gastric cancer and is a prerequisite for its management~\cite{wang2019chinese}.

Histopathological examinations begin by staining the sections with Hematoxylin and Eosin (H\&E), which are used to visualize the nuclei and cytoplasm of tissue sections, highlighting the fine structure of cells and tissues for physician observation~\cite{feldman2014tissue}.
The pathologist finds the diseased area by gross observation of the pathological slides with the naked eye. The pathologist then observes and diagnoses the diseased area of the pathological section using the low power microscope of the microscope. Pathologists can use high-power microscopes for careful observation and judgment.
For the entire pathological slice diagnosis process~\cite{tahiliani2021retrospective}, the following problems can be found: slice information is easy to ignore. This shows that there is subjectivity throughout the process.
The workload of pathologists is huge and the working hours are long, which is highly likely to lead to misdiagnosis.
Therefore, there is an urgent need to address the issues more intensively.

However, computer-aided diagnosis technology has advanced rapidly in recent years, and the emergence of medical image classification technology in computer vision technology can achieve fast and efficient help for doctors to examine gastric cancer tissue sections~\cite{nazarian2021diagnostic}.
Image classification techniques have brought new breakthroughs to discriminate benign and malignant cancer, distinguish between stages of tumor differentiation and differentiate tumor subtypes, as image classification techniques can provide valid information for pathologists to refer to during the diagnostic process~\cite{schmarje2021survey}.
In addition, the development direction of image classification technology is mainly to enhance the accuracy of classification algorithms and improve the anti-interference ability, ensemble learning becomes an effective solution, and it becomes especially important to find multiple efficient classification algorithms with complementarity properties~\cite{shinde2018review}.
Moreover, there is a lack of computer performance in practical work, and computer-aided medical image analysis often crops full-slice images into sub-size pictures~\cite{li2022hierarchical}.
Therefore, we compare the image classification performance of a large number of algorithms on sub-size images in order to expect to find algorithms with complementarity properties for ensemble learning to improve medical image classification performance.

The database used in this study is GasHisSDB~\cite{hu2022gashissdb}, containing 245,196 images, of which there are 97,076 abnormal images and 148,120 normal images. GasHisSDB is a database containing three sub-databases, including sub-database A (160$\times$160 pixels), Sub-database B (120$\times$120 pixels.), Sub-database C (80$\times$80 pixels). GasHisSDB provides the ability to distinguish between classical machine learning classifier performance and deep learning classifier performance~\cite{shin2016deep}. Details are given in Section 2.1.

Classical machine learning methods still have excellent classification results in the field of image classification~\cite{wang2021comparative}. Existing methods can extract different features of images and supply different performance of classifiers for image classification. Exploring different features using appropriate classifiers to obtain efficient classification results is the basis of using ensemble learning for medical images. Therefore, in this study, five different image features including two color features and three texture features are extracted for GasHisSDB. After extracting the features seven different classifiers are used for classification. Details are given in Section 2.2 and Section 2.3.

In the field of medical image classification, deep learning algorithms are the most effective algorithms, and Convolutional Neural Network (CNN) is a widely used model for image classification, which can extract information from original medical images and classify normal and abnormal case images~\cite{ker2017deep}. Recently, Visual Transformer, which were originally applied to Natural Language Processing tasks, have become popular in computer vision, and Vision Transformer (ViT) have effective classification results when trained on large amounts of data and can significantly reduce the computer hardware and software resources required for training~\cite{dai2021transmed}. CNN-based deep learning models, this study used VGG6, Inception-V3 and ResNet50. visual transformer-based deep learning models this study used VIT. the above four deep learning models use the same parameters with the same database: GasHisSDB. Details are given in Section 2.4.

This study makes the following contributions to the field of sub-size pathology image classification: 
\begin{itemize}
\itemsep=0pt
\item Extensive testing is done and the complementarity of different classification methods is found. 
\item According to the complementarity, it can provide a basis for future ensemble learning research.
\end{itemize}

This paper is structured as follows:In Section 2, we detail the dataset used, classical classification methods, and deep learning methods. In Section 3, we show the comparative experimental setup, evaluation metrics and experimental results.
In Section 4, we compare the experimental results and analyze them.
In Section 5, we summarize the research and suggest future research directions.
\section{Materials and methods}
\subsection{Datasets:GasHisSDB}

\begin{figure*}
	\centering
		\includegraphics[scale=.8]{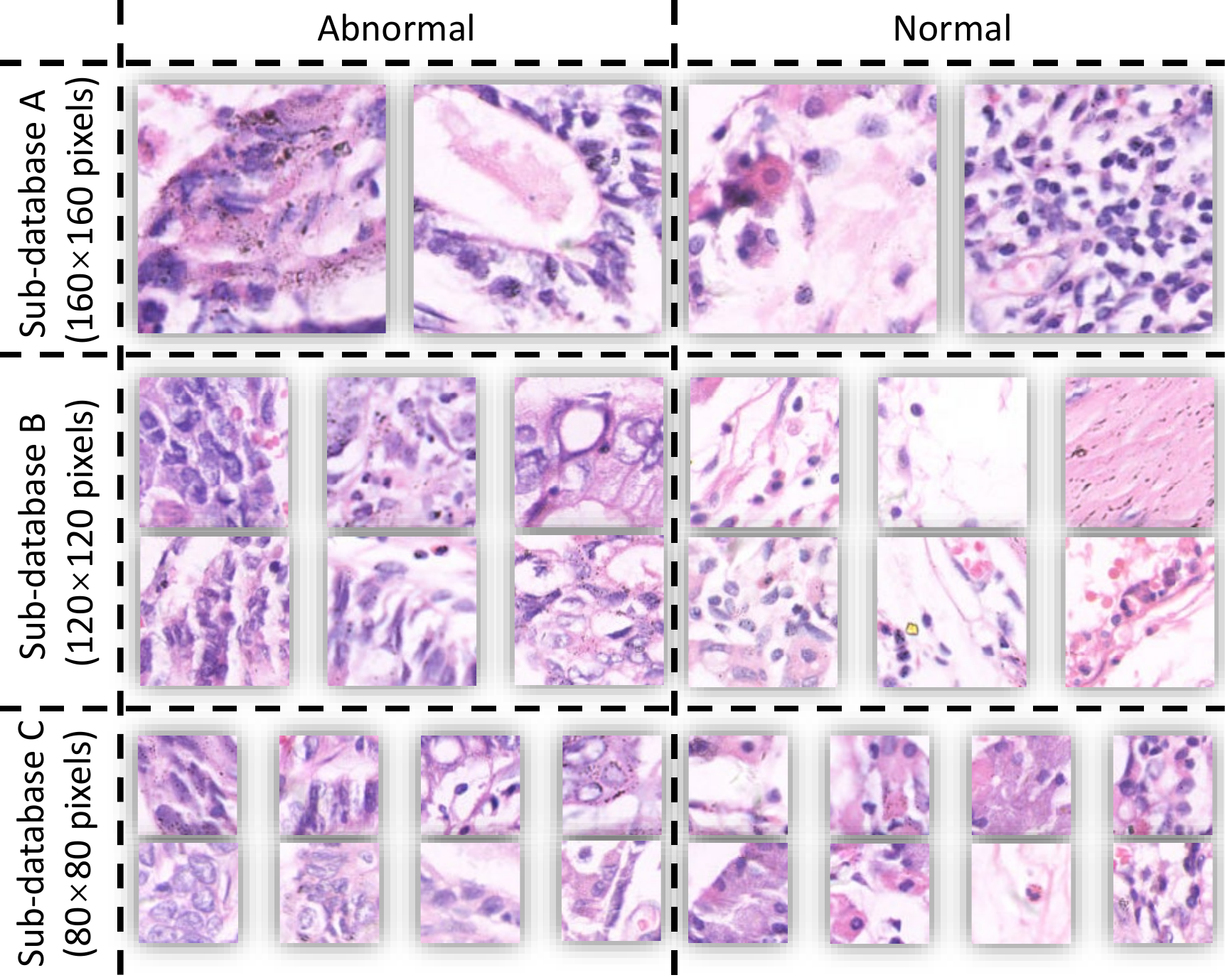}
		\caption{Example of GasHisSDB.}
	\label{FIG:1}
\end{figure*}

The publicly available dataset GasHisSDB is used in this study to compare the performance of various learning models, expecting to discover the complementarity of various models in ensemble learning~\cite{hu2022gashissdb}. The database contains three sub-datasets with a total of 245,196 images, and the size and number are shown in Table. \ref{tbl1}. The database is a sub-size gastric cancer pathology H\&E staining image database, which contains two categories of images: normal and abnormal. The abnormal image contains more than 50\% of the cancerous area, and the normal image is the image of the normal pathological slice tissue. Some examples of the GasHisSDB database are shown in Figure. \ref{FIG:1}.

\begin{table}[width=1\linewidth,cols=4,pos=h]
\caption{Dataset scale of GasHisSDB.}\label{tbl1}
\begin{tabular*}{\tblwidth}{@{} LLLLLLLL@{} }
\toprule
Sub-database name & Cropping size    & Abnormal & Normal  \\
\midrule
Sub-database A    & $160\times160$ pixels           & 13,124   & 20,160  \\
Sub-database B    & $120\times120$ pixels           & 24,801   & 40,460  \\
Sub-database C    & $80\times80$ pixels             & 59,151   & 87,500  \\
\bottomrule
Total             &                  & 97,076   & 148,120 \\
\bottomrule
\end{tabular*}
\end{table}

GasHisSDB contains images in png format acquired using electron microscopy. GasHisSDB contains two categories and the details of the two categories are shown below:
\begin{itemize}
\itemsep=0pt
\item Normal: each normal image does not contain cancerous regions. Each cell is almost free of anisotropy. In addition, the nuclei of the cells in the images have almost no mitosis and are arranged in a regular layer. Therefore, when observed under the light microscope, if no elimination of any cells and tissues is observed and the characteristics of a normal image are met, it can be judged as a normal image~\cite{japanese2011japanese}.
\item Abnormal: Each abnormal image contains more than 50\% of gastric cancer images. The general morphology of gastric cancer is mostly ulcerative. As the disease progresses, the cancer nest infiltrates from the mucosal layer to the muscular layer and plasma layer. The texture is hard and the cross-section is often grayish white. Under microscopic observation, the cancer cells can be arranged in nest-like, glandular vesicle-like, tubular or cord-like, and the boundary with the interstitium is usually clear. However, when cancer cells infiltrate into the stroma, the borders between them are not clear. Based on these facts, abnormal pathological images can be judged when cells are observed to form unevenly sized, irregularly shaped, and irregularly arranged glandular or adenoid structures~\cite{japanese2011japanese}.
\end{itemize}

\subsection{Methods of feature extraction}
To extract a variety of virtual features of GasHisSDB is a prerequisite for classification using classical machine learning classifiers.
In the comparison experiments, five methods are used to extract visual features from the database, including Color histogram, Luminance histogram, Histogram of Oriented Gradient (HOG), Local Binary Patterns (LBP) and Gray-level Co-occurrence Matrix (GLCM).

\subsubsection{Color histogram}
Among the different methods of feature extraction, the most common method to describe the color features of an image is color histogram.
The color histogram clearly represents the color spread in the image.
The color histogram has the characteristic of being unaffected by image rotation and shift changes and by further normalization of image scale changes.
It is especially applicable to describe images that are resistant to automatic segmentation and images that do not require consideration of the spatial location of subjects.
However, the color histogram does not characterize the partial spread of colors in an image, the spatial location of each color, and specific objects.
In this experiment, luminance histogram is used as the luminance feature. 
The luminance feature is expressed as a histogram of the average of the three color components.

\subsubsection{Texture features}

Texture as a visual feature of reflecting homogeneous phenomena in an image~\cite{humeau2019texture}.
That reflects the structure and arrangements of the surface structures on the surface of an object with slow or periodic changes.
A texture feature is not a pixel-based feature. It requires statistical computation of regions containing multiple pixels, such as the grayscale distribution of pixels and their surrounding spatial neighbors, and local texture information.
In addition, the global texture information is reflected as the repetition degree of local texture information.
In this experiment, three texture features are extracted, which are HOG, LBP and GLCM.

HOG is a feature descriptor commonly used in image processing for object detection.
Features are constructed by computing a histogram of the gradient direction of local regions of an image. 
HOG has the property of operating on the local units of the image. So it has the advantage of maintaining excellent invariance in terms of geometric and optical distortion of the image.
LBP has advantages such as gray invariance and rotation invariance, and the features are easy to compute.
GLCM is defined by the joint probability density of pixels at two locations, and is a second-order statistical feature about the variation of image brightness.
It not only reflects the distribution of luminance. It also reflects the distribution of positions between pixels with the same or similar luminance. The main statistical values are: Contrast, Correlation, Energy and Homogeneity.

\subsection{Classical classification models}

After the feature extraction step, complementarity comparison tests for image classification are performed using seven classical machine learning methods, including Linear Regression, $k$-Nearest Neighbor ($k$NN), naive Bayesian classifier, Random Forest (RF), linear Support Vector Machine (linear SVM), non-linear Support Vector Machine (non-linear SVM), and Artificial Neural Network (ANN).

Classical machine learning methods perform image classification by using virtual features. Linear Regression is a method to get a linear model as much as possible to accurately predict the real value output label. In Linear regression, least square function is used to establish the relationship between one or more independent variables~\cite{hope2020linear}. An easy and commonly used supervised learning method is $k$NN. The main idea of $k$NN is to first find the nearest $k$ samples based on the distance and then vote for the prediction result~\cite{guo2003knn}. The naive Bayesian classifier that is based on Bayesian decision theory in probability theory~\cite{yang2018implementation}. RF is a parallel integrated learning method based on a decision tree learner. RF adds random attribute selection to the training process of decision trees~\cite{shi2006unsupervised}. SVMs are divided into linear and nonlinear. The difference between the two is mainly that the kernel functions of both are different~\cite{suthaharan2016support}. Linear SVM maps training examples to points in space to maximize the gap between the two categories. Then, the new examples are mapped to the same space and predicted to belong to a category based on which side of the gap they fall on. In addition to performing linear classification, SVM can also use a kernel function to perform non-linear classification effectively, thereby implicitly mapping its input to a high-dimensional feature space. The ANN is a classification algorithm composed of a structure that simulates human brain neurons and is trained through a propagation algorithm~\cite{hopfield1988artificial}.

\subsection{Deep learning models}

Complementarity comparison experiments use deep learning models for classification of gastric cancer pathology images. First, the model is trained using training and validation sets generated from three sub-datasets of GasHisSDB. The test set is used in this experiment to evaluated the models' performance. Comparative analysis of multiple classification results is performed using the obtained evaluation metrics to determine if the classifiers would be complementarity in Ensemble learning. This experiment uses four deep learning models. Three of the models are based on CNNs, including VGG16, Inception-V3, and ResNet50. One more model corresponds to VT, which is ViT.

VGG is a convolutional neural network (CNN) improved by AlexNet, developed by Visual Geometry Group and Google DeepMind in 2014, and the most commonly used one in image classification is VGG16~\cite{simonyan2014very}.
In 2014, Google's InceptionNet made its debut at the ILSVRC competition. Several versions of InceptionNet have been developed, with Inception-V3 being one of the more representative versions of this large family~\cite{szegedy2016rethinking}.
Kaiming He et al. proposed ResNet to address the difficulty of training deep networks due to gradient disappearance. The most commonly used in the field of image classification is ResNet50~\cite{he2016deep}. 
In recent years, Alexey Dosovitskiy et al. have proposed the ViT model using transformer. This model is not only very effective in the field of natural language processing, but also provides good results in the field of image classification. Effectively reduces the dependence of computer vision on CNN~\cite{dosovitskiy2020image}.

\section{Experiment}

\subsection{Comparative experimental setup}

\begin{figure*}
	\centering
		\includegraphics[scale=.6]{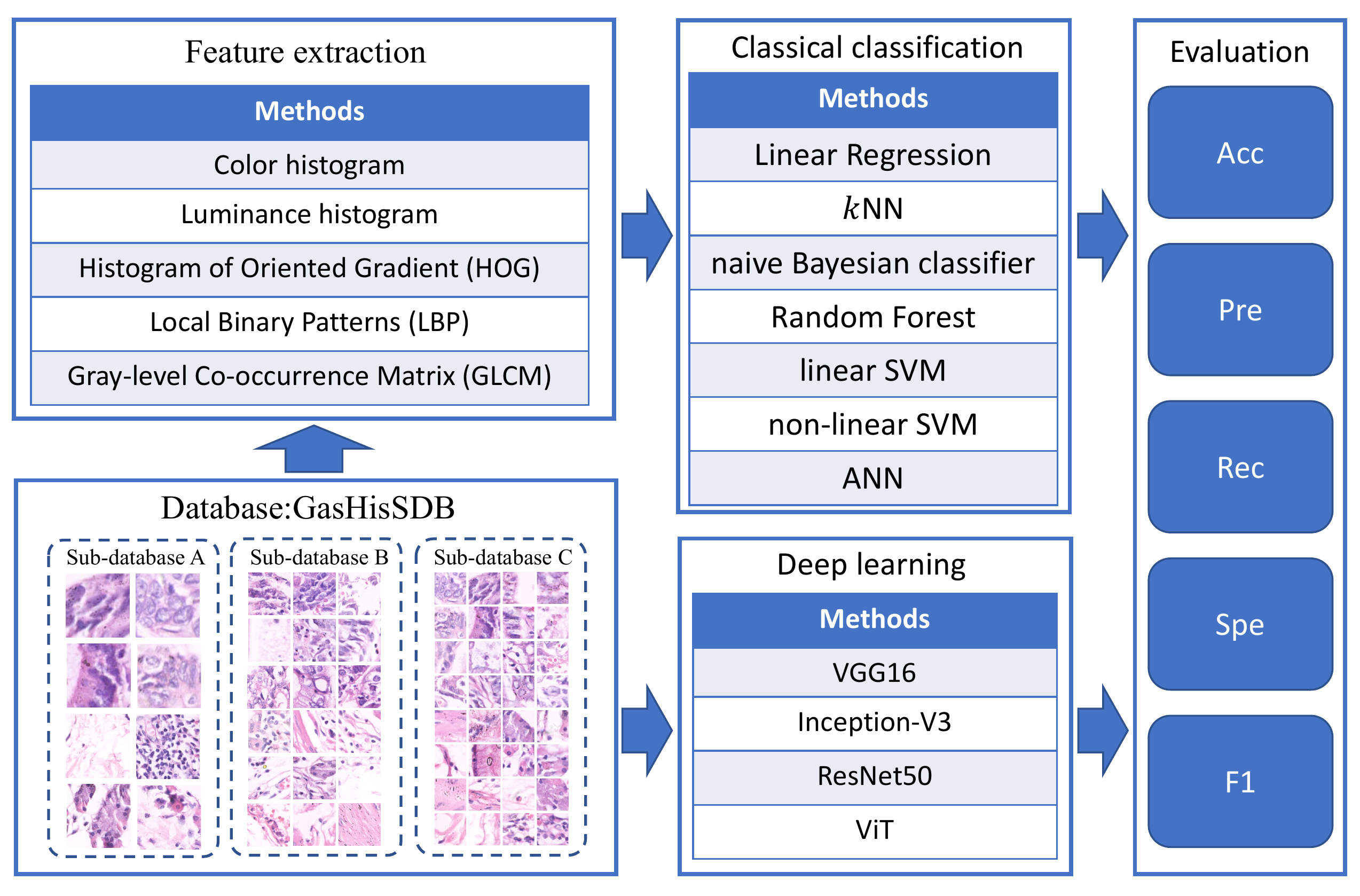}
		\caption{Flow chart of the complementarity comparison experiment.}
	\label{FIG:2}
\end{figure*}

The main process of complementarity experiments is divided into two parallel parts: The classification results of classical models and deep learning models are both analyzed and evaluated. The experimental flow is shown in Figure. \ref{FIG:2}.

The complementarity comparison experiment is conducted on a local computer with the Win10 operating system. the computer has 32 GB of running memory and is equipped with an 8 GB NVIDIA Quadro RTX 4000. 
In this experiment, the training set, validation set and test set are divided in the ratio of 4:4:2.

The classical programming software used for machine learning is Matlab R2020a (9.8.0.132 350 2). 
The different classical machine learning classification comparison experiments all used the same parameters.
The same parameters were used for all classification comparison experiments. In $k$NN, $k$ is set to 9. The number of trees in RF is set to 10. The kernel function of the nonlinear SVM is a Gaussian kernel. The ANN uses a 2-layer network with 10 nodes in the first layer and 3 nodes in the second layer. The number of epochs for ANN training is set to 500, the learning rate is set to 0.01, and the expected loss is set to 0.01.

The Pytorch version 1.7.1 framework in Deep Learning Python 3.6 is very mature, and the code for this part of the experiment is done using them.
This part of the experiment focuses on classifying GasHisSDB using four deep learning methods to observe model complementarity. A learning rate of 0.00002 is used for each model, and the batch size is set to 32. 100 epochs of experiments are performed to observe the classification results of this database on different models.

\subsection{Evaluation metrics}

The selection of evaluation indicators is important in complementarity comparison papers. In the experiments of this thesis, 
Accuracy (Acc) is the most significant metric, but also Precision (Pre), Recall (Rec), Specificity (Spe) and F1-score (F1) are selected. These selected metrics are very commonly used in comparison papers to analyze classifiers and thus better identify their complementarities to enhance and improve ensemble learning~\cite{liu2022aspect}.

In the case of positive-negative binary classification, true positives (TP) correspond to the number of positive samples that are accurately predicted. The number of negative samples predicted to be positive is called false positive (FP). The number of positive samples predicted to be negative samples is called false negative (FN). True Negative (TN) is the number of negative samples predicted accurately.

The five evaluation indicators are described below and the formulas are shown in Table. \ref{tbl6}.

\begin{enumerate}
\itemsep=0pt
\item Acc: Accuracy is the ratio of the number of correct predictions to the total number of samples.
\item Pre: Precision is a measure of accuracy, indicating the proportion of examples classified as positive that are actually positive.
\item Recl: Recall is a measure of coverage, a measure of the number of positive examples classified as positive examples, indicating the proportion of all positive examples classified as pairs, which measures the ability of the classifier to identify positive examples.
\item Spe: Specificity indicates the proportion of all negative cases that were scored correctly, and measures the classifier's ability to identify negative cases.
\item F1: F1-Score combines Precision and Recall. Accuracy is the ratio of the number of correct predictions to the total number of samples.
\end{enumerate}

\begin{table}[]
\caption{Evaluation metrics.}\label{tbl6}
\begin{tabular}{cc}
\hline
Assessment     & Formula               \\ \hline
Accuracy (Acc) & (TP+TN)/(TP+TN+FP+FN) \\
Precision (Pre)      & TP/TP+FP              \\
Recall (Rec)         & TP/TP+FN              \\
Specificity (Spe)    & TN/TN+FP              \\
F1-score (F1)      & 2$\times$(Pre$\times$Rec)/(Pre+Rec)         \\ \hline
\end{tabular}
\end{table}

\subsection{Experimental results}

We set up an experimental platform to conduct various classification experiments on three sub-databases of the GasHisSDB.
A large amount of experimental data is obtained for our experiments in order to investigate the complementarity of different methods.

The comparative results of classical machine learning methods are shown in Table \ref{tbl2}, Table. \ref{tbl3} and Table. \ref{tbl4}

Table. \ref{tbl5} show the comparison results of the deep learning methods.

\begin{table*}[width=2\linewidth,cols=11,pos=h]
\caption{Classification results of five image features using different classifiers in the 160$\times$160 pixels sub-database of GasHisSDB (In [\%]). The bold text in the table indicates the highest value of the classification result of different classifiers for the same feature.}\label{tbl2}
\resizebox{\textwidth}{65mm}{
\begin{tabular}{clrrrrrrrrrrrrr}
\hline
\multirow{2}{*}{\textbf{Freatures}}     & \multicolumn{1}{c}{\multirow{2}{*}{\textbf{Methods}}} & \multicolumn{1}{c}{\multirow{2}{*}{\textbf{Acc}}} & \multicolumn{1}{c}{\textbf{}} & \multicolumn{1}{l}{\textbf{}} & \multicolumn{4}{c}{\textbf{Abnormal}}                                                                                                                           & \multicolumn{1}{c}{\textbf{}} & \multicolumn{1}{l}{\textbf{}} & \multicolumn{4}{c}{\textbf{Normal}}                                                                                                                             \\ \cline{6-9} \cline{12-15} 
                                        & \multicolumn{1}{c}{}                                  & \multicolumn{1}{c}{}                              & \multicolumn{1}{c}{\textbf{}} & \multicolumn{1}{l}{\textbf{}} & \multicolumn{1}{l}{\textbf{Pre}} & \multicolumn{1}{l}{\textbf{Rec}} & \multicolumn{1}{l}{\textbf{Spe}} & \multicolumn{1}{l}{\textbf{F1}} & \multicolumn{1}{l}{\textbf{}} & \multicolumn{1}{l}{\textbf{}} & \multicolumn{1}{l}{\textbf{Pre}} & \multicolumn{1}{l}{\textbf{Rec}} & \multicolumn{1}{l}{\textbf{Spe}} & \multicolumn{1}{l}{\textbf{F1}} \\ \hline
\multirow{6}{*}{Color histogram}        & LR                                                    & 83.29                                             &                               &                               & 81.32                                  & 80.42                               & 85.54                                    & 80.87                                 &                               &                               & 84.80                                  & 85.54                               & 80.42                                    & 85.17                                 \\
\hline
\end{tabular}}
\end{table*}

\begin{table*}[width=2\linewidth,cols=11,pos=h]
\caption{Classification results of five image features using different classifiers in the 120$\times$120 pixels sub-database of GasHisSDB (In [\%]). The bold text in the table indicates the highest value of the classification result of different classifiers for the same feature.}\label{tbl3}
\resizebox{\textwidth}{65mm}{
\begin{tabular}{clrrrrrrrrrrrrr}
\hline
\multirow{2}{*}{\textbf{Freatures}}     & \multicolumn{1}{c}{\multirow{2}{*}{\textbf{Methods}}} & \multicolumn{1}{c}{\multirow{2}{*}{\textbf{Acc}}} & \multicolumn{1}{c}{\textbf{}} & \multicolumn{1}{l}{\textbf{}} & \multicolumn{4}{c}{\textbf{Abnormal}}                                                                                                                           & \multicolumn{1}{c}{\textbf{}} & \multicolumn{1}{l}{\textbf{}} & \multicolumn{4}{c}{\textbf{Normal}}                                                                                                                             \\ \cline{6-9} \cline{12-15} 
                                        & \multicolumn{1}{c}{}                                  & \multicolumn{1}{c}{}                              & \multicolumn{1}{c}{\textbf{}} & \multicolumn{1}{l}{\textbf{}} & \multicolumn{1}{l}{\textbf{Pre}} & \multicolumn{1}{l}{\textbf{Rec}} & \multicolumn{1}{l}{\textbf{Spe}} & \multicolumn{1}{l}{\textbf{F1}} & \multicolumn{1}{l}{\textbf{}} & \multicolumn{1}{l}{\textbf{}} & \multicolumn{1}{l}{\textbf{Pre}} & \multicolumn{1}{l}{\textbf{Rec}} & \multicolumn{1}{l}{\textbf{Spe}} & \multicolumn{1}{l}{\textbf{F1}} \\ \hline
\multirow{6}{*}{Color histogram}        & LR                                                    & 83.46                                             &                               &                               & 80.01                                  & 75.28                               & 88.47                                    & 77.57                                 &                               &                               & 85.38                                  & 88.47                               & 75.28                                    & 86.90                                 \\
\hline
\end{tabular}}
\end{table*}

\begin{table*}[width=2\linewidth,cols=11,pos=h]
\caption{Classification results of five image features using different classifiers in the 80$\times$80 pixels sub-database of GasHisSDB (In [\%]). The bold text in the table indicates the highest value of the classification result of different classifiers for the same feature.}\label{tbl4}
\resizebox{\textwidth}{65mm}{
\begin{tabular}{clrrrrrrrrrrrrr}
\hline
\multirow{2}{*}{\textbf{Freatures}}     & \multicolumn{1}{c}{\multirow{2}{*}{\textbf{Methods}}} & \multicolumn{1}{c}{\multirow{2}{*}{\textbf{Acc}}} & \multicolumn{1}{c}{\textbf{}} & \multicolumn{1}{l}{\textbf{}} & \multicolumn{4}{c}{\textbf{Abnormal}}                                                                                                                           & \multicolumn{1}{c}{\textbf{}} & \multicolumn{1}{l}{\textbf{}} & \multicolumn{4}{c}{\textbf{Normal}}                                                                                                                             \\ \cline{6-9} \cline{12-15} 
                                        & \multicolumn{1}{c}{}                                  & \multicolumn{1}{c}{}                              & \multicolumn{1}{c}{\textbf{}} & \multicolumn{1}{l}{\textbf{}} & \multicolumn{1}{l}{\textbf{Pre}} & \multicolumn{1}{l}{\textbf{Rec}} & \multicolumn{1}{l}{\textbf{Spe}} & \multicolumn{1}{l}{\textbf{F1}} & \multicolumn{1}{l}{\textbf{}} & \multicolumn{1}{l}{\textbf{}} & \multicolumn{1}{l}{\textbf{Pre}} & \multicolumn{1}{l}{\textbf{Rec}} & \multicolumn{1}{l}{\textbf{Spe}} & \multicolumn{1}{l}{\textbf{F1}} \\ \hline
\multirow{6}{*}{Color histogram}        & LR                                                    & 82.22                                             &                               &                               & 78.17                                  & 77.59                               & 85.35                                    & 77.88                                 &                               &                               & 84.93                                  & 85.35                               & 77.59                                    & 85.14                                 \\
\hline
\end{tabular}}
\end{table*}

\begin{table*}[width=2\linewidth,cols=11,pos=h]
\caption{Classification results of four deep learning classifiers on GasHisSDB (In [\%]). The bold text in the table indicates the maximum value or the best index of the classification results of different categories.}\label{tbl5}
\resizebox{\textwidth}{65mm}{
%
} & \textbf{Acc}                    & \multicolumn{1}{c}{\textbf{Category}} & \multicolumn{1}{c}{\textbf{Pre}} & \multicolumn{1}{c}{\textbf{Rec}} & \multicolumn{1}{c}{\textbf{Spe}} & \multicolumn{1}{c}{\textbf{F1}} \\ \hline
\multirow{10}{*}{$160\times160$ pixels}                                                & \multirow{2}{*}{VGG16}        & \multirow{2}{*}{100}                                                 & \multirow{2}{*}{268.16}                                           & \multirow{2}{*}{100}                                          & \multirow{2}{*}{13873}                                              & \multirow{2}{*}{95.90}          & Abnormal                              & 93.8                                   & 96.0                                & 95.9                                     & 94.9                                  \\
                                                                     &                               &                                                                      &                                                                   &                                                               &                                                                     &                                 & Normal                                & 97.3                                   & 95.9                                & 96.0                                     & 96.6                                  \\
                                                                     & \multirow{2}{*}{Inception-V3} & \multirow{2}{*}{100}                                                 & \multirow{2}{*}{89.69}                                            & \multirow{2}{*}{92}                                           & \multirow{2}{*}{10296}                                              & \multirow{2}{*}{94.57}          & Abnormal                              & 94.1                                   & 92.0                                & 96.2                                     & 93.0                                  \\
                                                                     &                               &                                                                      &                                                                   &                                                               &                                                                     &                                 & Normal                                & 94.9                                   & 96.2                                & 92.0                                     & 95.5                                  \\
                                                                     & \multirow{2}{*}{ResNet50}     & \multirow{2}{*}{100}                                                 & \multirow{2}{*}{83.12}                                            & \multirow{2}{*}{84}                                           & \multirow{2}{*}{10023}                                              & \multirow{2}{*}{\textbf{96.09}} & Abnormal                              & 94.6                                   & 95.6                                & 96.4                                     & 95.1                                  \\
                                                                     &                               &                                                                      &                                                                   &                                                               &                                                                     &                                 & Normal                                & 97.1                                   & 96.4                                & 95.6                                     & 96.7                                  \\
                                                                     & \multirow{4}{*}{ViT}          & \multirow{2}{*}{100}                                                 & \multirow{2}{*}{31.17}                                            & \multirow{2}{*}{97}                                           & \multirow{2}{*}{2587}                                               & \multirow{2}{*}{86.21}          & Abnormal                              & 83.8                                   & 80.6                                & 89.9                                     & 82.2                                  \\
                                                                     &                               &                                                                      &                                                                   &                                                               &                                                                     &                                 & Normal                                & 87.7                                   & 89.9                                & 80.6                                     & 88.8                                  \\
                                                                     &                               & \multirow{2}{*}{400}                                                 & \multirow{2}{*}{31.17}                                            & \multirow{2}{*}{399}                                          & \multirow{2}{*}{10014}                                              & \multirow{2}{*}{92.23}          & Abnormal                              & 92.1                                   & 87.8                                & 95.1                                     & 89.9                                  \\
                                                                     &                               &                                                                      &                                                                   &                                                               &                                                                     &                                 & Normal                                & 92.3                                   & 95.1                                & 87.8                                     & 93.7                                  \\ \hline
\multirow{10}{*}{$120\times120$ pixels}                                                & \multirow{2}{*}{VGG16}        & \multirow{2}{*}{100}                                                 & \multirow{2}{*}{268.16}                                           & \multirow{2}{*}{100}                                          & \multirow{2}{*}{26105}                                              & \multirow{2}{*}{\textbf{96.47}} & Abnormal                              & 96.7                                   & 94.0                                & 98.0                                     & 95.3                                  \\
                                                                     &                               &                                                                      &                                                                   &                                                               &                                                                     &                                 & Normal                                & 96.4                                   & 98.0                                & 94.0                                     & 97.2                                  \\
                                                                     & \multirow{2}{*}{Inception-V3} & \multirow{2}{*}{100}                                                 & \multirow{2}{*}{89.69}                                            & \multirow{2}{*}{98}                                           & \multirow{2}{*}{19719}                                              & \multirow{2}{*}{95.83}          & Abnormal                              & 94.6                                   & 94.4                                & 96.7                                     & 94.5                                  \\
                                                                     &                               &                                                                      &                                                                   &                                                               &                                                                     &                                 & Normal                                & 96.6                                   & 96.7                                & 94.4                                     & 96.6                                  \\
                                                                     & \multirow{2}{*}{ResNet50}     & \multirow{2}{*}{100}                                                 & \multirow{2}{*}{83.12}                                            & \multirow{2}{*}{94}                                           & \multirow{2}{*}{19087}                                              & \multirow{2}{*}{95.94}          & Abnormal                              & 96.2                                   & 93.0                                & 97.8                                     & 94.6                                  \\
                                                                     &                               &                                                                      &                                                                   &                                                               &                                                                     &                                 & Normal                                & 95.8                                   & 97.8                                & 93.0                                     & 96.8                                  \\
                                                                     & \multirow{4}{*}{ViT}          & \multirow{2}{*}{100}                                                 & \multirow{2}{*}{31.17}                                            & \multirow{2}{*}{100}                                          & \multirow{2}{*}{4077}                                               & \multirow{2}{*}{89.44}          & Abnormal                              & 87.0                                   & 84.9                                & 92.2                                     & 85.9                                  \\
                                                                     &                               &                                                                      &                                                                   &                                                               &                                                                     &                                 & Normal                                & 90.9                                   & 92.2                                & 84.9                                     & 91.5                                  \\
                                                                     &                               & \multirow{2}{*}{500}                                                 & \multirow{2}{*}{31.17}                                            & \multirow{2}{*}{496}                                          & \multirow{2}{*}{20410}                                              & \multirow{2}{*}{94.59}          & Abnormal                              & 93.5                                   & 93.4                                & 95.3                                     & 93.2                                  \\
                                                                     &                               &                                                                      &                                                                   &                                                               &                                                                     &                                 & Normal                                & 95.4                                   & 95.9                                & 92.5                                     & 95.6                                  \\ \hline
\multirow{10}{*}{$80\times80$ pixels}                                                 & \multirow{2}{*}{VGG16}        & \multirow{2}{*}{100}                                                 & \multirow{2}{*}{268.16}                                           & \multirow{2}{*}{90}                                           & \multirow{2}{*}{62152}                                              & \multirow{2}{*}{\textbf{96.12}} & Abnormal                              & 94.2                                   & 96.3                                & 96.0                                     & 95.2                                  \\
                                                                     &                               &                                                                      &                                                                   &                                                               &                                                                     &                                 & Normal                                & 97.4                                   & 96.0                                & 96.3                                     & 96.7                                  \\
                                                                     & \multirow{2}{*}{Inception-V3} & \multirow{2}{*}{100}                                                 & \multirow{2}{*}{89.69}                                            & \multirow{2}{*}{99}                                           & \multirow{2}{*}{43926}                                              & \multirow{2}{*}{95.41}          & Abnormal                              & 95.5                                   & 93.0                                & 97.0                                     & 94.2                                  \\
                                                                     &                               &                                                                      &                                                                   &                                                               &                                                                     &                                 & Normal                                & 95.3                                   & 97.0                                & 93.0                                     & 96.1                                  \\
                                                                     & \multirow{2}{*}{ResNet50}     & \multirow{2}{*}{100}                                                 & \multirow{2}{*}{83.12}                                            & \multirow{2}{*}{97}                                           & \multirow{2}{*}{41992}                                              & \multirow{2}{*}{96.09}          & Abnormal                              & 96.2                                   & 94.0                                & 97.5                                     & 95.1                                  \\
                                                                     &                               &                                                                      &                                                                   &                                                               &                                                                     &                                 & Normal                                & 96.0                                   & 97.5                                & 94.0                                     & 96.7                                  \\
                                                                     & \multirow{4}{*}{ViT}          & \multirow{2}{*}{100}                                                 & \multirow{2}{*}{31.17}                                            & \multirow{2}{*}{89}                                           & \multirow{2}{*}{8247}                                               & \multirow{2}{*}{90.23}          & Abnormal                              & 86.3                                   & 90.1                                & 90.3                                     & 88.2                                  \\
                                                                     &                               &                                                                      &                                                                   &                                                               &                                                                     &                                 & Normal                                & 93.1                                   & 90.3                                & 90.1                                     & 91.7                                  \\
                                                                     &                               & \multirow{2}{*}{500}                                                 & \multirow{2}{*}{31.17}                                            & \multirow{2}{*}{496}                                          & \multirow{2}{*}{41135}                                              & \multirow{2}{*}{94.57}          & Abnormal                              & 93.1                                   & 93.4                                & 95.3                                     & 93.2                                  \\
                                                                     &                               &                                                                      &                                                                   &                                                               &                                                                     &                                 & Normal                                & 95.6                                   & 95.3                                & 93.4                                     & 95.4                                  \\ \hline
\end{tabular}}
\end{table*}

\section{Evaluation of results}

\subsection{Evaluation of classical machine learning methods}

\subsubsection{On $160\times160$ pixels sub-database}

This section focuses on the classification results of classical 
machine learning methods for the 160$\times$160 sub-database.

The color histogram has the highest number of items among all features.
According to Tab. \ref{tbl2}, the classical machine learning classifier on the color histogram, the best performer is RF with an accuracy of 85.99\%.
In addition, in color histogram, the classification accuracy of the three classifiers reached around 80\%, which are LR, $k$NN and ANN.
All SVM classifiers perform poorly on color histogram features.
However, color histogram on GasHisSDB, the naive Bayesian classifier, cannot get the classification effect because of the existence of a large number of low luminance statistics with zero values in the three color channels.

The luminance as the mean value of the color and its histogram as a feature does not gain a better classification accuracy.
Because of this, luminance histogram also has the above problem on the naive Bayesian classifier. 
The classification results of the naive Bayesian classifier for these two color features are therefore not presented in the Table. \ref{tbl2}. 
RF shows robustness in two features and obtains the highest accuracy rate of 79.13\% using luminance histogram for classification.
However, the LR, $k$NN and ANN classifiers that perform better on color histogram significantly drop on luminance histogram.

The classification effect of HOG on all classifiers is not very effective and the accuracy is very close. The difference is not much distributed between 53\%-62\%.

On the contrary, the distribution of LBP image classification accuracy is particularly scattered, with the highest Linear Regression classifier reaching 74.29\%, followed by ANN reaching 71.84\%. The lowest linear SVM classification effect is less than 50\%.

The classification effect of the four statistic values of GLCM is 71.39\% only for RF, and other classifiers are also above 60\%.
It is worth noting that the accuracy of non-linear SVM with other features except color histogram and GLCM  has not changed at all, which is 60.58\%. The accuracy of non-linear SVM classifier with color histogram is 56.09\% and the accuracy of GLCM's non-linear SVM classifier is 67.76\%.

\subsubsection{On $120\times120$ pixels sub-database}

Here, we focus on the comparison of the experimental results of the $120\times120$ pixels sub-database.
The experimental results are shown in Table. \ref{tbl3}.
In general, compared with $160\times160$ pixels sub-database classification results, $120\times120$ pixels sub-database classification results except for color histogram, the rest of the best classifiers remain unchanged.

The four better-performing classifiers on color histogram feature still perform better, and the accuracy rate fluctuates slightly, resulting in the $k$NN classifier reaching the best accuracy rate of 86.32\%.
The classification performance of the two SVM classifiers on the features of color histogram is still not ideal.
Naive Bayesian classifier is still not suitable for color histogram and luminance histogram features.
The linear SVM effect of luminance histogram classifier has been greatly improved in the classification of the $120\times120$ pixels sub-database.
The accuracy of other classifiers on the features of luminance histogram has little change.
The HOG feature still does not perform well in every classifier.
The highest accuracy rate is only 62.35\% of ANN.
The classification results of LBP and GLCM features are similar to the classification effect on the $160\times160$ pixels sub-database. The best accuracy rate on LBP is a linear regression with a precision rate of 73.34\%. The best accuracy rate on GLCM is that the RF reaches 71.15\%.
Similarly, the non-linear SVM of $120\times120$ pixels sub-database also has the problem of constant accuracy of multiple features.

\subsubsection{On $80\times80$ pixels sub-database}

The classification results of the $80\times80$ pixels sub-database are shown in Table. \ref{tbl4}.
The overall best classifier on each feature remains the same as that of the best classifier for each feature corresponding to the $120\times120$ pixels sub-database except for HOG features that have a small gap between each classifier.

Compared with the classification results of the other two sub-databases, the classification effect of each classifier on color histogram and luminance histogram has no particularly large fluctuations.
It confirms the consistency of the three databases of GasHisSDB.

The classification accuracy of color histogram is still polarized. The four excellent classifiers reach about 80\%, and the other two are about 60\%.
RF still shows robustness in the task of luminance histogram classification and is the best classifier with an accuracy of 75.10\%.
The classification accuracy distribution of HOG features is denser than that of the other two sub-databases.
The highest is only 59.87\%.
Due to the reduced sample size, each classifier has different degrees of accuracy reduction in addition to the naive Bayesian classifier for LBP features and GLCM features.
The best classifier for LBP feature is still linear regression which reaches 70.92\%.
The highest accuracy rate of LBP feature has become 68.84\% of $k$NN.
In the classification results of the $80\times80$ pixels sub-database, the naive Bayesian classifier of color histogram and luminance histogram is not applicable, and, except for the GLCM feature, the problem that the accuracy of the nonlinear SVM classifier does not change still exists.

\subsection{Evaluation of deep learning methods}

\subsubsection{On $160\times160$ pixels sub-database}

According to Table. \ref{tbl5}, on $160\times160$ pixels sub-database, all deep learning models have better classification results than classical machine learning methods.
The VGG model with the longest training time and the largest model size has an accuracy above 95\%.
Inception-V3 and ResNet50 have better model size and training time than VGG16. However, Inception-V3 has lower accuracy than VGG16, and ResNet50 has the highest accuracy of 96.09\%, which is the highest among all models.
ViT is a Transformer-based classifier with an accuracy of 86.21\%, but still higher than the classification accuracy of all traditional machine learning methods on this sub-database.
Significantly, ViT achieves such accuracy with only 1/4 of the training time and 1/3 of the model size compared to ResNet. Also, the accuracy curve is still trending upward and the loss function is still not fully converged.

\subsubsection{On $120\times120$ pixels sub-database}

According to the Table. \ref{tbl5}, the classification results are excellent on the sub-database of $120\times120$ pixels.
Due to the large number of training samples, VGG16 is the classifier with the highest accuracy of 96.47\% on this sub-database.
However, the training time is doubled compared to that on the $160\times160$ sub-database.
The accuracies of 95.83\% and 95.94\% are obtained for Inception-V3 and ResNet50, respectively.
Due to the increase in the amount of training data, ViT also gained an accuracy improvement, rising to 89.44\%.

\subsubsection{On $80\times80$ pixels sub-database}

According to Table. \ref{tbl5}, the classification results of the $80\times80$ subdatabase can be seen. It is the sub-database with the largest number of samples, and the accuracy of the four classifiers only changes slightly. 
VGG16 performs stably with an accuracy of 96.12\%, which is the classification model with the highest accuracy.
The lowest accuracy is still the ViT model with the least training time, at 90.23
It is worth noting that the training time of ViT is 13.26\% of that of the highest accurate VGG16 on this sub-database.

\subsection{Additional experiment}

As stated in Section 4.2.1, ViT did not converge completely within 100 epochs.
Experiments are added in this section to explore the performance of ViT, and the results are reflected in the last row of each sub-database in Table. \ref{tbl5}.
The same parameter conditions were maintained for all additional experiments.
In the additional experiments for the $160\times160$ sub-database, the control training time was similar to that of Inception-V3 and ResNet running 100 epochs. ViT runs 400 epochs and the accuracy reaches 92.23\%.
In the other two sub-databases with larger amount of data, again when controlling for the same training time as Inception-V3 and RseNet50. At this time, the accuracy of ViT models for the $120\times120$ pixel sub-database and the $80\times80$ pixel sub-database improves to 94.59\% and 94.57\%, respectively.
The model size of ViT has a great advantage. Moreover, these image classification results reach the general level of medical image classification.

\subsection{Discussion}

This chapter compares the classification results of different classifiers from the Linear Regression to Visual Transformer on the $160\times160$ pixels, $120\times120$ pixels and $80\times80$ pixels sub-databases of the GasHisSDB.
The classification performance of each method on GasHisSDB reflects complementarity.

Classical machine learning methods have a rigorous theoretical foundation. Their simplified ideas can show good classification results on some specific features and algorithms.

This experimental platform shows that seven classifiers for GLCM classification on three sub-databases with little difference in accuracy, where the naive Bayesian classifier has significantly higher Rec than Spe for the abnormal category, and the linear SVM has slightly higher Rec than Spe.
It shows that these two classifiers are better in classifying the abnormal category.
However, the Spe of the other classification models are higher than the Rec, indicating that they are more effective in classifying the normal category.
The same phenomenon occurs for every feature of every sub-database. There exist classifiers with high Rec values or high Spe values in the same condition. Such a result can be a powerful indication of the existence of this complementarity of these classifiers.

However, deep learning methods are still far ahead of classical machine learning methods in terms of image classification accuracy and experiment workload.

The evaluation metrics for deep learning models are generally high, but complementarity in the field of machine learning also occurs in the field of deep learning.
For example, the Spe of Inception-V3 and ResNet50 on sub-database C for abnormal category classification is high, but the high Rec of VGG16 can be well performed to the complementarity of the above two models.

The selection of suitable classifiers is the primary problem of ensemble learning, and after relevant experiments in the complementarity comparison experimental platform, it can be observed that these classifiers exhibit different performances. The complementarity possessed by these classifiers can adequately meet the needs of ensemble learning.

\section{Conclusion and futures works}

In practice, machine performance often limits model training for large size images, and finding multiple classification models with complementarity types is the basis for ensemble learning. For sub-sized images, this experiment tries a large number of classification models to find their complementarity and thus improve the efficiency of ensemble learning.

The experimental results show that complementarity in machine learning does exist for different classifiers of the same feature. Different classifiers for the same feature include classifiers that classify the abnormal category well and classifiers that classify the normal category well. This is a powerful indication of the complementarity among classifiers.

The evaluation metrics of the deep learning models are both very excellent.
There are models that are less effective in classifying the abnormal category than the normal category.
In this case, selecting the appropriate model that performs well for the abnormal category can contribute to ensemble learning.  Complementarity can also be demonstrated in this situation.

There are still many excellent methods that have not been added to the experimental platform.
Moreover, the recently popular ViT excels in the field of image processing, but ViT does not show significant experimental results on sub-size images.
In the future, we will add more models to explore the complementarity nature of ensemble learning on sub-size images to improve the efficiency of ensemble learning.

\section*{Acknowledgements}
This work is supported by the "National Natural Science Foundation of China" (No. 61806047). 
We also thank Miss. Zixian Li and Mr. Guoxian Li for their important discussion in this work.

\section*{Declaration of competing interest}

The authors declare that they have no conflict of interest.

\bibliographystyle{unsrt}
\bibliography{ref}

\end{document}